\documentclass[a4paper,11pt]{article}
\usepackage{pos}

\title{Astro-photography as an effective tool for Outreach and Education: IACT in exposition}
\ShortTitle{Astro-photography as an effective tool for Outreach and Education}

\author*[a,b,c]{Simone Iovenitti}
\author[b,c]{Chiara Righi}
\author[c]{Stefano Orsenigo}
\author[c]{Riccardo Sgarro}

\affiliation[a]{Universit\'a degli Studi di Milano, Dipartimento di Fisica\\
Via Giovanni Celoria 16, 20133 Milano (MI), Iatly}

\affiliation[b]{INAF - Osservatorio Astronomico di Brera,\\
Via E Bianchi 46, 23807 Merate (LC), Italy}

\affiliation[c]{Associazione culturale PhysicalPub,\\
Via A. Vivaldi 8, 20054 Segrate (MI), Italy}

\emailAdd{simone.iovenitti@inaf.it}

\abstract{In our epoch, images are a powerful way to convey a message to a large audience. Through the use of amazing astronomical photographs, science can be communicated effectively at different levels, to a very diverse audience of all ages. In fact, astrophotography combines aesthetic appeal with the illustration of the science behind astronomical phenomena. This is the aim of the exhibit “A che Punto è la NOTTE - A scientific exhibition of astrophotography” organized by us in Italy, in October 2020, with the partnership of the cultural association PhysicalPub. Many different authors, both single individuals and professional or amateur observatories, were asked to send their best pictures. The 54 astronomical images chosen by a scientific committee, categorised in three different topics (night landscape, deep sky, instrumentation), were seen by more than 2000 visitors and 11 school groups (despite the difficult period due to the COVID pandemic). A free audio-guide, available on-line through a web-application developed on purpose, delivered scientific explanations of images for self-guided tours. Conferences and guided tours were also organized.
The highlight of the exhibit were four mirrors from the MAGIC telescope and an ASTRI scale-model that allowed an in-depth description of how an Imaging Atmospheric Cherenkov Telescope (IACT) works, introducing the science of VHE cosmic radiation.
We will summarize the main difficulties in organizing this event and the feedback we had from the visitors. The exhibit is still available online, visiting the website \url{mostrascientifica.it} or via the web audio-guide (english and italian) at \url{guida.mostrascientifica.it}.}

\FullConference{37$^{\rm{th}}$ International Cosmic Ray Conference (ICRC 2021)\\
		July 12th -- 23rd, 2021\\
		Online -- Berlin, Germany}

\begin{document}
\maketitle

\section{Introduction}      \label{section:1}
\noindent
Our everyday life is based on technology. Actually, the whole society we live in would not be the same without the most important technological achievements of the last decades. Basically, they come from scientific research and hence, in order to further improve the current condition, today the world is oriented to the development of scientific disciplines and people are always more interested in science and in the work of scientists. Curiosity is surely an evident intrinsic characteristic of every human being, if there are the conditions to express it. Unfortunately, it is not always easy to satisfy this natural instinct, as very often science requires many years of hard studies to be understood, and sometimes also very uncommon skills. However, there are cases where complicated processes or exotic phenomena can be grasped by the mind as represented in a very simple way: an image. Our life is driven by images (advertisement, social media, and so on), so why don’t we use them also to convey scientific knowledge? They are an effective medium, immediate, very attractive to the public, and accessible to everyone. For all these reasons we decided to develop a project for outreach and educational purposes, in the field of astronomy and astrophysics, based on the usage of amazing pictures taken with the particular technique of astrophotography. In this contribution we describe the whole project and our goals (section~\ref{section:2}), its implementation (section~\ref{section:3}) and the scientific content (section~\ref{section:4}) that spans all the fields of astrophysics with particular attention to the very-high-energy gamma-ray astronomy and to Cherenkov telescopes.
We had extremely positive feedback by the audience, as it is reported in the conclusion (section~\ref{section:6}), and hence we believe that our experience constitutes an example of how positive can be a huge outreach event for the communication of astronomy and in general for awakening the enthusiasm for science.\\
The entire project was developed thanks to the Italian cultural association PhysicalPub \cite{php}, that carries out a great effort to spread the culture of astronomy and the passion for science, believing that it will not only improve our lives, but also the condition of the whole planet.
 
\begin{figure}
    \centering
    \includegraphics[width=0.9\textwidth]{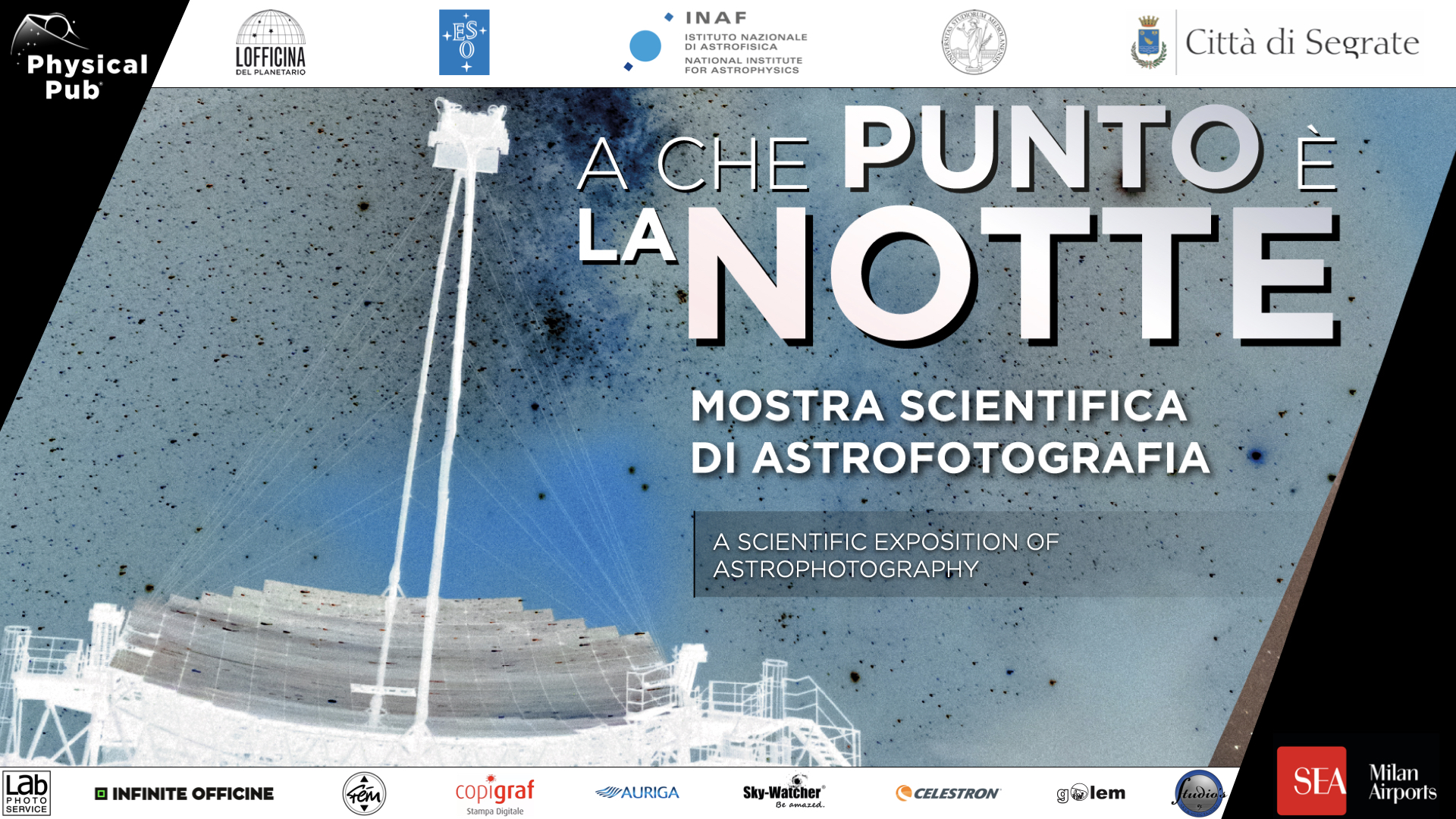}
    \caption{Main graphics of the exhibition: the original photo (by C.Righi) shows the MAGIC telescope.}
    \label{fig:1}
\end{figure}

\section{The project}      \label{section:2}
\noindent
In order to communicate the beauty of the Universe together with the passion for science, astrophotography is a very effective tool: it is a highly specialized technique of photography, providing amazing images of the sky representing lots of particular physics phenomena \cite{astro},\cite{astro2}. For this reason, at the end of 2019, we started to work on a new project for outreach and education based on a big exposition of astrophotography, where every picture is connected to a specific astrophysics topic. The name of this project was “A che Punto \'e la NOTTE?”, which means: “what is left of the NIGHT?”. It is an extract from the Bible, but it has nothing to do with religion in our project. Instead, we intended this sentence as a question about the state of the art in astrophysics, our current knowledge of the universe: we have reached considerable results but there is still a lot of research to do.\\
We had two main goals in our project. The first one is to communicate fascinating astrophysics topics to a very diverse audience, in order to spread the culture for astronomy and the passion for science. The second one is to promote people and places where the interest for sky observation is grown and spread. Observatories, associations, societies of passionate astronomers, both professional and amateur: we wanted to create a situation where people can meet these realities, keep in touch with them, so to take part in their activities maybe in the future. While pursuing these goals, we got another surprising result in addition: our exhibition worked as a connection bridge between different entities doing very similar activities, which have never met each other before. We were very happy to favor this process, as we deeply believe that science is, first of all, collaboration.

\subsection{A large collaboration}       \label{section:2.1}
\noindent
To realize our project we were supported by several partners and sponsors (their logos are shown in figure \ref{fig:1}). In fact, several established institutions encourages our idea: the European Southern Observatory (ESO), the University of Milan (UNIMI), the Italian National Institute of Astrophysics (INAF), the LOfficina del Planetario (Planetarium of Milan) and others\footnote{ ~A complete list of sponsors and partners can be found on our website \cite{php}.
}. The City of Segrate (a small town in Italy, close to Milan) gave us the necessary logistic support, while we were funded by SEA (Milan Airports) and GoLem (research group of the INAF-Osservatorio Astronomico di Brera, Merate section) to cover all the expenses: thanks to our sponsors, the exhibition and all the side activities were always completely free for the public.\\
To collect the pictures for the exhibition we opened a call on our website \cite{php} and in a few months we had a surprising answer: more than 200 photographs were sent on behalf of professional astrophotographers, scientific collaborations, and astronomical observatories, both professional and amateur. It was really hard to choose “only” 54 pictures for the exhibition: in the selection we did not consider only the beauty of the shots, but also their scientific content. The complete list of the exhibitors can be found on our website \cite{php} and in the slides of this conference. From the point of view of contributions, this exhibition was the largest one in Italy in this field.

\subsubsection{Scientific commission}      \label{section:2.2}
\noindent
We had a scientific board that supervised the accuracy of all the scientific contents of this project. The commission is composed of two full professors of the Universit\'a degli Studi di Milano, physics department (N. Ludwig, A. Mennella), a researcher of the Italian National Institute for Astrophysics (A. Wolter) and an expert of science communication from the Planetarium of Milan (A. Cassetti). They reviewed all the scientific descriptions of the pictures (both the short ones for the captions and the long ones for the audio-guide) and took part in the selections of the photographs. Moreover, they choose the invited speakers for a cycle of 5 conferences that we proposed as a side activity during the opening period (see section \ref{section:3.1}).
\begin{figure}
    \centering
    \includegraphics[width=0.8\textwidth]{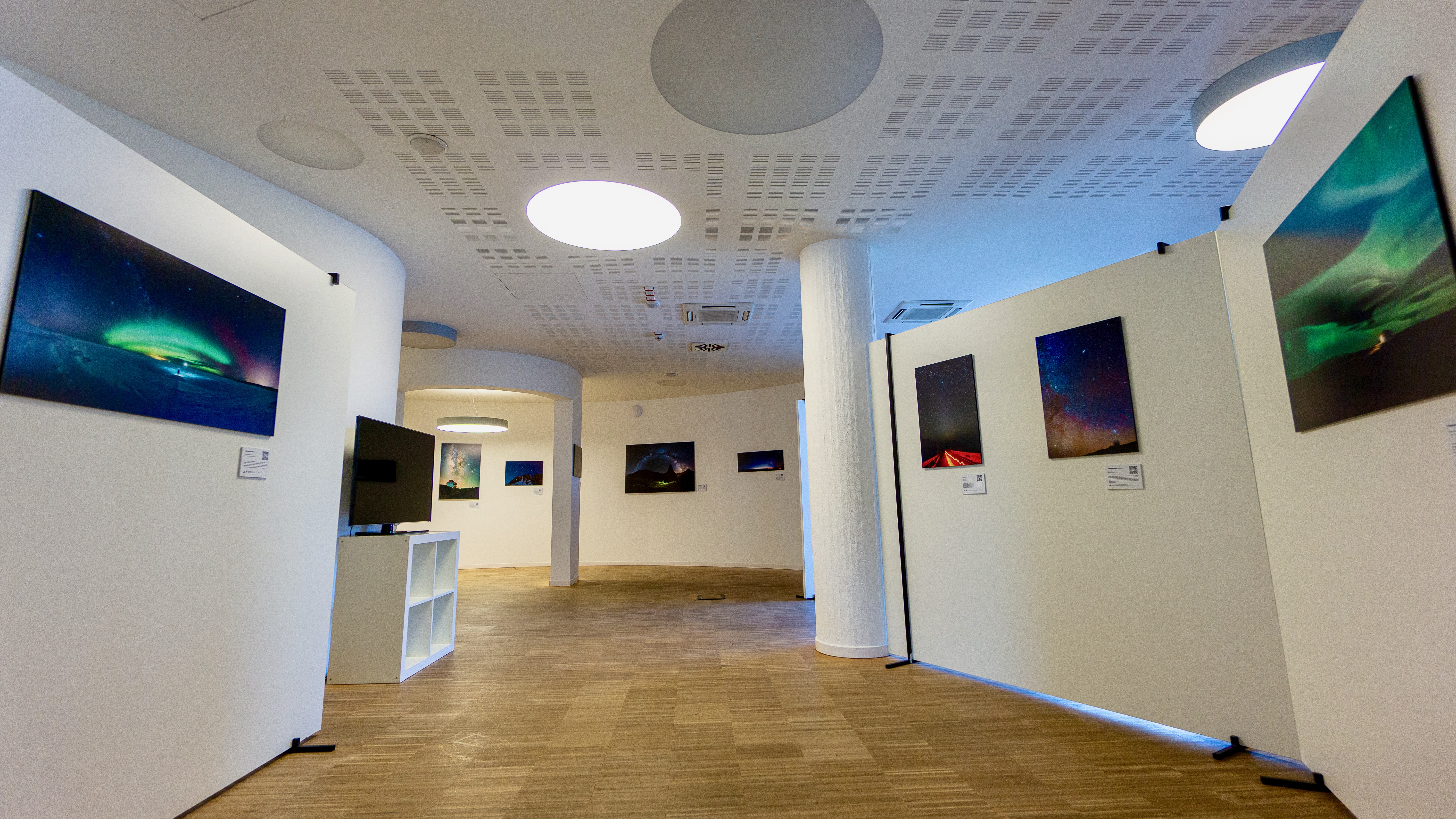}
    \caption{The first part of the exhibition in Segrate: the ``night landscape'' section.}
    \label{fig:Segrate}
\end{figure}

\section{Staging of the exhibition}      \label{section:3}
\noindent
We concretized our project at the end of 2020, in Italy, in two venues: Segrate (from 12 Sept. to 18 Oct) and Genova (from 22 Oct to 1 Nov.), where we inserted our initiative into the context of the national Festival of Science. The 54 photographs that we selected for the exhibition were printed in large formats on photographic paper and then applied on rigid panels (PIUMA). We arranged the pictures on the walls of the exhibition area (see figure \ref{fig:Segrate}), divided in three sections: night landscape, deep sky and instrumentation (see section \ref{section:4}). Next to each picture we hung a small plate in A6 format in rigid material (FOREX) containing the title, the author, the details of the shot, the astronomical or geographical position and a brief description of the scientific topic related to the photograph. To have further information, on each plate there was a QR-code pointing at the web audio-guide that we developed (see section \ref{section:5}), allowing the visitors to listen to long and accurate explanations while standing in front of the picture they are interested in.
Actually, the web audio-guide is the most important feature of our project: it constitutes a remote access to our exhibition, making it a virtual event available for free, at any time.\\
The staging of the exhibition was completed with 7 giant photos (1.2 m x 1.2 m) from ESO, partner of our project, that we arranged in the entrance of the exhibition to welcome the visitors. A lot of info-graphic material was also available (brochures of sponsors, flyers of the observatories, posters and panels) while at the exit there was a lot of free merchandise (pens, stickers, postcards and bookmarks).

\subsection{Side events}      \label{section:3.1}
\noindent
The exhibition was completed by a series of side events, during the opening period.
A cycle of 5 scientific lectures on different topics for the general audience was organized into a beautiful auditorium in Segrate. These conferences were available in-presence, but they were also broadcasted in live streaming on YouTube, where they are still available today \cite{php}. 
On every Friday night the exhibition was open until 11 p.m. and guided tours for the general public were organized both for groups and single visitors. Only for school groups, the guided tours were available also during the morning, to allow the teachers to visit the exhibition with their students during the lesson time. Moreover, for school groups we prepared a contest: each class had to present a little project regarding some astronomical observations to be carried out in the context of the school program. Our scientific commission examined all the proposals and the best one was awarded a real telescope!
The last side activity was a very immersive experience: a journey into a virtual environment created on purpose, containing some of the astronomical objects represented in the pictures, with the use of VR viewers. Unfortunately, because of the covid-19 pandemic, we had to eliminate this activity from the visit to our exhibition.
\begin{figure}
    \centering
    \includegraphics[width=\textwidth]{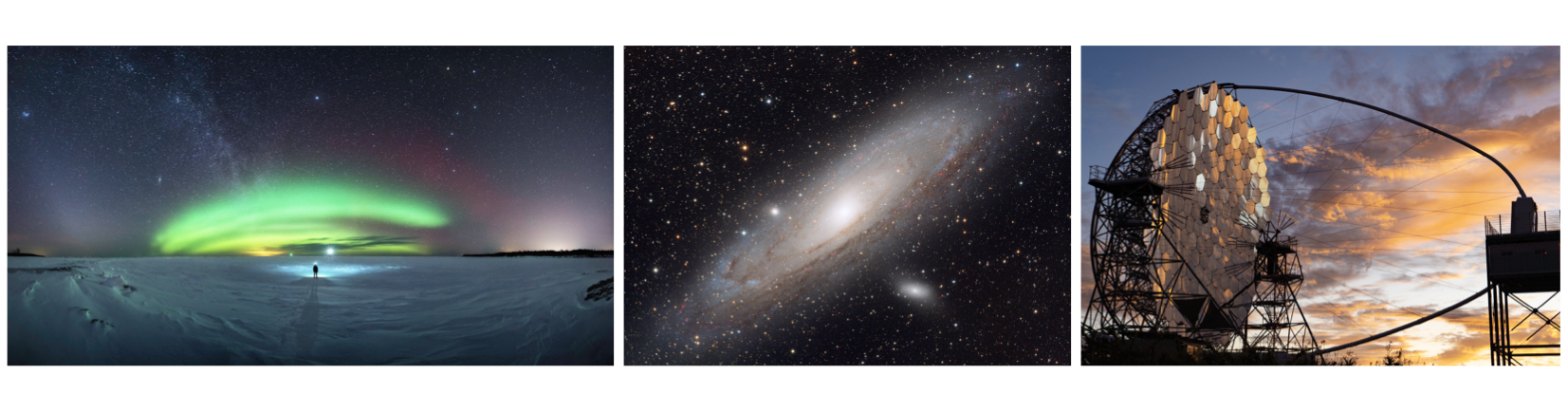}
    \caption{An example of a photo for each section of the exhibition. \textit{Left:} a Northern Light in the ``Night landscape'' section.
    \textit{Center:} Andromeda galaxy in the ``Deep-sky section''.
    \textit{Right:} the Large Size Telescope (LST, a prototype for the CTA observatory) in the ``Instrumentation'' section.}
    \label{fig:images}
\end{figure}

\section{Scientific content}      \label{section:4}
 \noindent
As the main goal of our project was the communication of science, every single picture in the exhibition was associated with a specific astrophysics topic. A very brief description of each argument was provided in the captions (the plates next to the pictures), while detailed and advanced explanations were available both in the audio-guide and during the guided tours, both for school groups and the general public. On these occasions, the guide focused the attention of the audience on only a few pictures, following an itinerary chosen by the audience among a selection of possible thematic routes, such as “colors in astronomy”, “the Solar System”, “high-energy astrophysics”, and others \cite{mostra}. The guided tours went across the three main sections into which the exhibition is divided, that we briefly describe hereafter, while some examples are reported in figure \ref{fig:images}.

\begin{figure}
    \centering
    \includegraphics[width=0.7\textwidth]{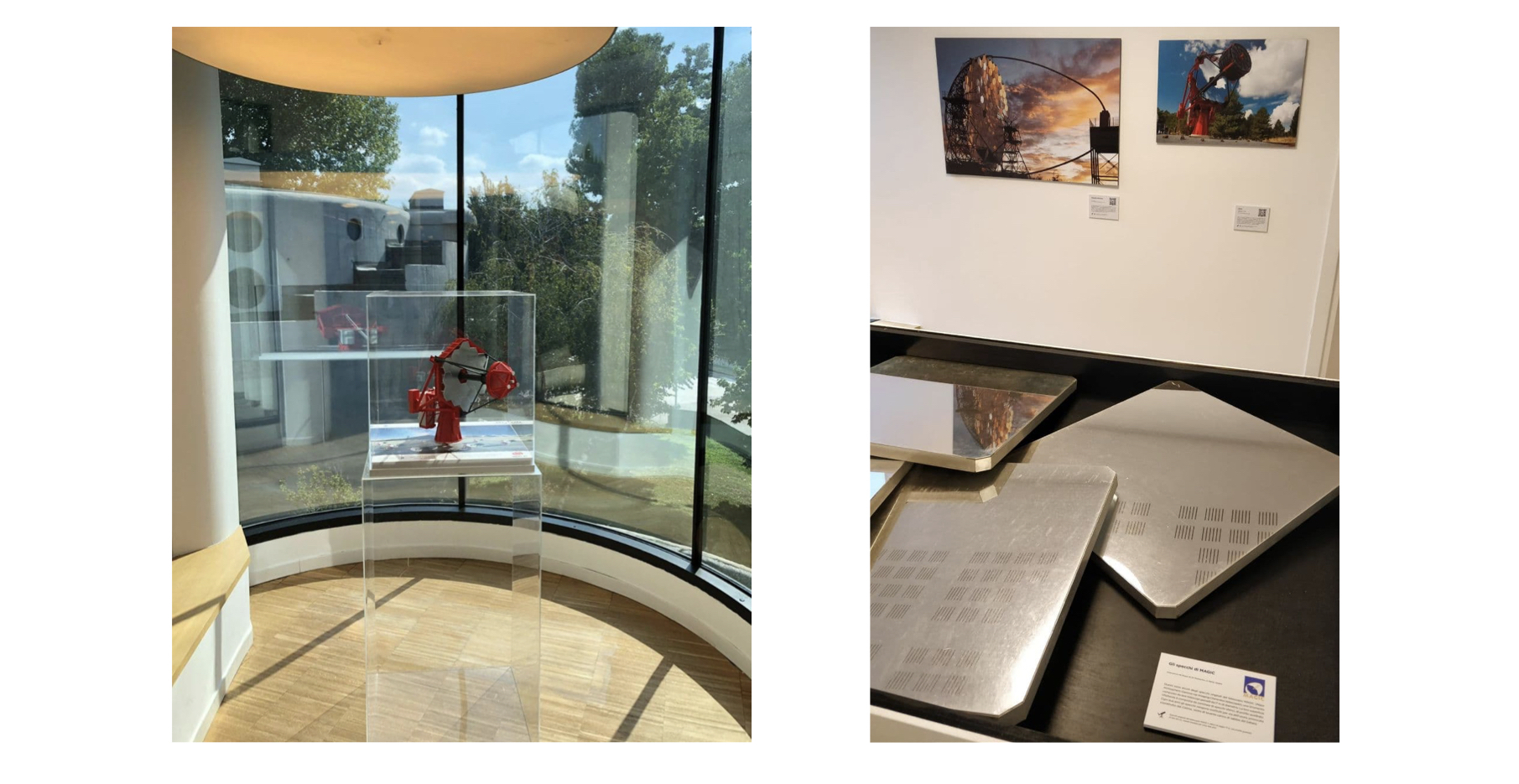}
    \caption{Non-photographic material related to Cherenkov astronomy: a scale-model of the ASTRI telescope \textit{(left)} and four original mirrors from the MAGIC-I telescope \textit{(right)}.}
    \label{fig:mirrors}
\end{figure}

\begin{itemize}
  \item[] \textit{Night landscape.} Wide-field night photographs that allowed us to focus on comets, meteors, zodiacal light, but also on atmospheric phenomena, such as Northern lights, air-glow, light pollution, and others. Specific photograph techniques, like the startrails or the panorama, were also discussed in this section.

  \item[] \textit{Deep-sky.} Images taken with a telephoto lens or a telescope provide a clear view on nebulae, galaxies, clusters but also on the planets of the Solar System and the Moon.
In this section we focused also on the difficulties in realizing these pictures: filters, exposure time, aberrations and others.

    \item[] \textit{Instrumentation.} Various pictures, taken mostly by researchers, representing the protagonists of astronomical research: telescopes, detectors, mirrors and men at work.
\end{itemize}
The range of topics covered by our exhibition spans all the macro-areas of astrophysics, but particular attention was dedicated to the field of Cherenkov astronomy, as some of us are directly involved in the MAGIC collaboration \cite{MAGIC} and the ASTRI project \cite{ASTRI}. In particular, the operation of Imaging Atmospheric Cherenkov Telescopes (IACTs) was described in detail, together with the principles of the stereoscopic view, and also the very high-energy (VHE) gamma-ray astronomy was explained to the public. To this end, we exploited not only the pictures of the exhibitions, but also two special objects: a scale-model of the ASTRI telescope, realized in additive manufacturing, and four original mirrors (50 cm x 50 cm) of the MAGIC-I telescope, that visitors could touch and inspect (figure \ref{fig:mirrors}).

\section{Web audio-guide developed on purpose}      \label{section:5}
\noindent
To allow single visitors to appreciate the scientific content of every picture, we provided a completely free web audio-guide, available online using the web browser of any device, thanks to a platform that we developed on purpose.
The audio-guide was designed in order to let people listen to the scientific explanation using their own mobile devices, equipped with earphones, while standing in front of the pictures. With this strategy, personal tablets or smartphones become the audio-guides, avoiding sharing material among visitors, in accordance with the prescriptions to reduce the risks associated with the COVID-19 pandemic.
To access the audio-guide we put a big QR-code at the beginning of the exhibition, but also smaller QR-codes were present in every caption, pointing exactly at the description of that specific picture. Inside the web application, it is possible to navigate between a description and another, so that the QR-code must be scanned only once. Moreover, the audio-guide is available also remotely, outside the WiFi area of the exhibition, so that everyone can re-access the scientific content everywhere, in every moment.

\begin{figure}
    \centering
    \includegraphics[width=0.9\textwidth]{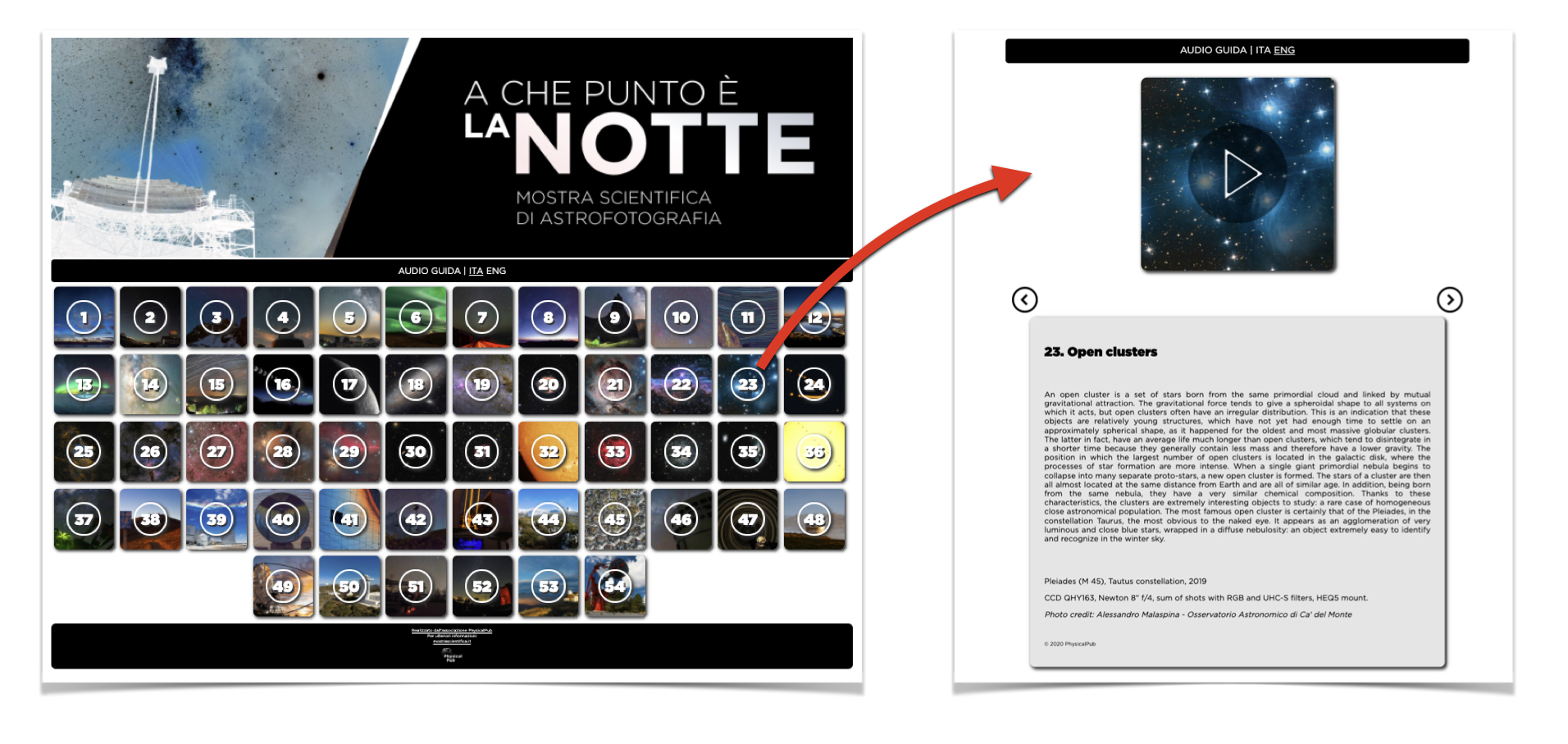}
    \caption{Home page of the audio-guide (left) and an example of a page related to a specific picture (the number 23) with all the details and the audio file (right).}
    \label{fig:guide}
\end{figure}

\subsection{Realization and structure}      \label{section:5.1}
\noindent
The audio-guide consisted of a web application with a back-end coded in Node JS, while the front end was written in ELM language and transpiled to JavaScript, with the page style managed by CSS. The application was hosted on a web server (a virtual private server, VPS), publicly exposed on the internet using Apache and accessible using an easy url \cite{guidamostra}. The same configuration of the VPS was replicated on an local server, completely independent, made with a RaspberryPi 4 and a WiFi router, so that the audio-guide was available in the area of the exhibition also without an internet connection (in case of low signal or simply not to consume the mobile data of the users).
The structure of the web application was composed of a main page with all the thumbnails of the pictures and their number (figure \ref{fig:guide}, \textit{left}). By clicking on any picture, the user can open a dedicated page, containing the details of the shot\footnote{ ~Similar to an EXIF file.} and the instrumentation adopted, the astronomical or geographical coordinates, the credits to the author and a long text with the complete scientific description of the topic. On the top of the page there is a preview of the picture and a button to reproduce the audio file (figure \ref{fig:guide}, \textit{right}).
The audio files were recorded in a professional studio, with the help of a sound technician and a producer. The audio tracks contain the voice of two professional speakers, a man and a woman, who read aloud the detailed scientific descriptions that we wrote. The audio recording corresponds to the long text which is present in the web application, so that the visitors can access both the written and the audio version.
The whole audio-guide is available both in Italian and in English, even if at present the audio recording is available only in Italian.

\section{Conclusion}      \label{section:6}
\noindent
Despite the difficulties related to the COVID-19 pandemic, our exhibition constitutes a success under several points of view. First, we had very excellent feedback from the audience: more than 2000 visitors saw our exhibition and lots of people left positive comments both at the venue and on social media. This is a hint that people were really pleased and sincerely interested in the science that we explained to them. Also teachers reported that their students often continued to elaborate on our topics, and this is the most important result of our initiative: to awaken the curiosity and the passion for science. Moreover, among the visitors we saw especially young people, and lots of undergraduate students. This is probably because we made a massive usage of social media for promoting our event (Instagram pictures, Facebook events and so on) but also because of the communication strategy that we adopted: accurate scientific explanations are more attractive to the general public, when conveyed with astonishing astronomical images. For this reason, we can say that astrophotography, as we adopted it, is definitely validated as a very effective outreach and educational tool.

\section*{Acknowledgments}
\noindent
We would like to thank all the people who gave their energy, their time and their creativity for the realization of our project. A complete list can be found on the audio-guide web page \cite{grazie}, but in particular we would like to thank all the members of the association Physicalpub, whose help was really the key element in this experience. Moreover, we would like to thank the ASTRI project and the MAGIC Collaboration, and in particular Marina Manganaro, for her constant support and the corrections in this manuscript.


\begin{thebibliography}{99}

\bibitem{exhibit}
Farrona, A. \& Vilar, Rocio. 2016. Nuclear and Particle Physics Proceedings. 273-275. 1225-1228. 10.1016/j.nuclphysbps.2015.09.194. 

\bibitem{astro}
De Leo Winkler, Mario \& Canalizo, Gabriela \& Wilson, Gillian. 2016. International Journal of STEM Education. 3. 10.1186/s40594-016-0053-0. 

\bibitem{astro2}
Sparks \& Kruse, 2019, 2019ASPC..524..115S, ASPC, 524, 115

\bibitem{php}
\url{https://physical.pub}

\bibitem{mostra}
\url{https://mostrascientifica.it}

\bibitem{guidamostra}
\url{https://guida.mostrascientifica.it}

\bibitem{grazie}
\url{https://guida.mostrascientifica.it/ringraziamenti}

\bibitem{MAGIC}
Aleksic J., et al., 2016,  Astroparticle Physics, 72, 76

\bibitem{ASTRI}
Pareschi, G. 2016, Proc. SPIE, 9906, 99065T


\end{thebibliography}
\end{document}